\documentclass[10pt,twocolumn,twoside]{IEEEtran}
\IEEEoverridecommandlockouts
\usepackage{cite}
\usepackage{amsmath,amssymb,amsfonts}
\usepackage{algorithmic}
\usepackage{graphicx}
\usepackage{textcomp}
\usepackage{xcolor}
\usepackage{cite}
\usepackage{amsmath,amssymb,amsfonts}
\usepackage{algorithmic}
\usepackage{textcomp}
\usepackage{xcolor}
\usepackage{algorithm}
\usepackage{comment}
\usepackage{setspace}
\usepackage{color}
\usepackage{booktabs}
\usepackage{multicol}
\usepackage{multirow}
\usepackage{epstopdf}
\usepackage{subfigure}

\def\BibTeX{{\rm B\kern-.05em{\sc i\kern-.025em b}\kern-.08em
    T\kern-.1667em\lower.7ex\hbox{E}\kern-.125emX}}
\begin{document}
\title{Minimizing Age-of-Information for Fog Computing-supported Vehicular Networks with Deep Q-learning}
\author{\IEEEauthorblockN{Maohong~Chen\IEEEauthorrefmark{1}, Yong~Xiao\IEEEauthorrefmark{1},  Qiang~Li\IEEEauthorrefmark{1}, Kwang-cheng Chen\IEEEauthorrefmark{2} \\
		\IEEEauthorblockA{\IEEEauthorrefmark{1}School of Electronic Information and Communications, Huazhong Univ. of Science \& Technology, China}\\
		\IEEEauthorblockA{\IEEEauthorrefmark{2} Department of Electrical Engineering, University of South Florida, FL}
}}
\maketitle

\begin{abstract}
Connected vehicular network is one of the key enablers for next generation cloud/fog-supported autonomous driving vehicles. Most connected vehicular applications require frequent status updates and Age of Information (AoI) is a more relevant metric to evaluate the performance of wireless links between vehicles and cloud/fog servers. This paper introduces a novel proactive and data-driven approach to optimize the driving route with a main objective of guaranteeing the confidence of AoI. In particular, we report a study on three month measurements of a multi-vehicle campus shuttle system connected to cloud/fog servers via a commercial LTE network. We establish empirical models for AoI in connected vehicles and investigate the impact of major factors on the performance of AoI. We also propose a Deep Q-Learning Netwrok (DQN)-based algorithm to decide the optimal driving route for each connected vehicle with maximized confidence level. Numerical results show that the proposed approach can lead to a significant improvement on the AoI confidence for various types of services supported.
\end{abstract}


\section{Introduction}
Connected vehicular system has been envisioned as one of the key enablers critical for next generation cloud/fog-supported autonomous driving vehicles. Compared to the traditional self-driving vehicular systems which mostly rely on the on-board computer for driving behavior processing and decision making, connected vehicles can receive real-time information from the Internet as well as other vehicles periodically or on-demand for improved safety and driving experience. This enables a multitude of novel applications such as real-time situation awareness, coordinated lane changing, and joint path planning for congestion avoidance. Connected vehicles can also upload computationally intensive tasks to the cloud data center (CDC) for intelligent driving decisions such as path planning, vulnerable road user discovery, traffic monitoring and early warning of potential dangerous events\cite{ge2015spatial}. 

Many connected vehicle-enabled applications require frequent status updating and the updated information must be successfully delivered to the targeted servers or vehicles in a timely manner. A relevant metric for quantifying the timeliness and freshness of the information/status delivery is Age-of-Information (AoI) which is defined as the time elapsed since the last received update of status information has been generated by the source \cite{5984917}. Guaranteeing a tolerable AoI in safety-related event broadcasting and information exchanging is of critical importance for vehicular networking systems.

One way to reduce AoI is to allow connected vehicles to be supported by fog computing. Fog computing network in a holistic framework for supporting computationally intensive applications that require low latency, high reliability, local context-awareness, and security. It consists of a large number of low-cost and often decentralized mini-servers performing computing, storage, and control closer to end users (i.e., connected vehicles). Fog computing has been commonly recognized as one of the most important components for future smart vehicular systems. To accelerate its adoption and commercial application, the European Telecommunications Standards Institute (ETSI) has formed an Industry Specifications Group (ISG) to standardize fog computing \cite{hu2015mobile}. OpenFog \cite{openfog2017openfog}, an industry-academia consortium, has also been created to promote standardization and dissemination of fog computing. Fog computing has also been embraced by mobile network operators (MNOs) as a way to create new business opportunities, increase revenues, and reduce costs. Major MNOs, including AT\&T, Deutsche Telekom, and China Mobile, have announced plans to integrate fog computing into their network infrastructure to support emerging services such as robotic manufacturing, autonomous cars, and augmented/virtual reality (AR/VR) applications. Due to its cost and space limitation compared to CDC, fog computing is not intended to replace the cloud data center but to complement it. In particular, fog nodes can handle localized low-latency context-aware services, while relying on remote large-scale CDCs to support more computationally intensive but latency-tolerant services.

Despite its great promise\cite{ge2016user}, cloud/fog computing-supported vehicular system is being hindered by several challenges. First of all, fog nodes are typically resource-limited devices. How to dynamically distribute users' workload between CDC and fog nodes according to the supported services and the corresponding service requirement is a very challenging task. Second, AoI of the wireless connection between vehicles and fog nodes can exhibit strong temporal and spatial variation. Finally,  AoI is also closely related to the driving behavior such as driving speed and routes.





In this paper, we introduce a novel proactive and data-driven approach for optimizing the driving route with the main objective to guarantee the confidence of AoI. We begin our work with an empirical analysis of AoI recorded on a vehicular system. Particularly we report a three-month measurement campaign on a multi-vehicle campus shuttle system connected with fog/cloud server through a commercially available LTE network with over 1,000,000 number of samples. We evaluate the impact of three main factors: update frequency, choice of cloud and fog servers, and processing delay at fog/cloud servers, on the empirical PDFs of AoI in a vehicular system. We observe that compared to other elements such as time of measurement and driving speed, the location diversity plays a major role in the statistics of AoI. In addition, although the instantaneous AoI value varies substantially from second to second at each given location, the empirical PDF of AoIs in each specific location can be considered as stationary. By evaluating and comparing empirical PDF recorded at different location. We observe that the PDF of AoIs exhibits strong spatial correlation. We then introduce a modified K-means-based approach to classify the empirical PDF of AoI measured at different locations throughout our considered area into a limited number of distribution function each of which corresponds to a set of regions that experience similar empirical PDFs of AoI. We investigate the driving route planning problem for the campus shuttles to maximize the confidence level of AoI when driving through different routes between the starting and finish points. We observe that since each fog node can only be connected with a limited number of vehicles at each time and therefore, the driving routes of all the vehicles must be carefully planned to avoid network congestion as well as traffic overload at some fog nodes coverage area. To address this problem, we propose a Deep Q-Network (DQN)-based algorithm for deciding the optimal driving routes for each vehicle with maximized confidence level during the entire driving route. Extensive simulations have been conducted and numerical results show a significant improvement on the AoI confidence for various types of connected vehicular services supported.
\vspace{-0.1cm}
\section{Related Works}

{\bf AoI of Connected Systems:} Most existing works focused on optimizing the resource allocation schemes for wireless networks to minimize AoI. In particular, the authors in \cite{sun2017update} studied the relationship between AoI and queueing delay and found that 
increasing the waiting time between consecutive data transmission will improve the AoI performance. In \cite{kam2015effect}, the authors studied the tradeoff between the network resource utilization and the instantaneous AoI and analyzed the transmission policy under various situations. In this paper, we focus on the performance of AoI in connected vehicular systems. We studied various factors that can influence AoI of connected vehicles when driving into different locations.  

{\bf Fog Computing-based Vehicular Systems:} Fog computing is a promising solution to reduce the service latency, especially for computationally intensive services\cite{xiao2018ti}. Therefore, most existing works focused on how to improve the utilization efficiency of the limited computational  resources at fog nodes. For example, the authors in \cite{8824922} proposed a threshold-based policy for dynamically switching between cloud and fog nodes to minimize the latency when driving. In \cite{7457059}, the authors considered a model with three policies to allocate fog resource for each task. A policy for improving the quality of experience by applying fog computing was established in \cite{mahmud2019quality}. In this paper, We consider the AoI in a hybrid system consisting of both cloud and fog. We evaluate the AoI performance supported by  
cloud and fog based on a long-term measurement study on a commercially available LTE network.

{\bf DQN for Dynamic Vehicular Systems:} DQN has been widely used in solving wireless communication problems. For example, the authors in \cite{8666109} proposed a network slicing approach to dynamically allocate computing resources with DQN. In \cite{8422586}, the authors applied DQN to find an optimal transmission policy that minimizes the communication cost. In \cite{8486393}, the authors indicated that there exists a temporal connectivity in a traffic network. In this paper, we consider a path-planning model to dynamically schedule the driving route with DQN.


\vspace{-0.1cm}
\section{Architecture Overview}
A cloud/fog computing-supported vehicular network consists of the following elements:
\begin{itemize}
	\item[1)] Vehicles: a set of $K$ connected vehicles ${\cal K}=\left\{1,\ldots ,K\right\}$ each of which drives from one location to another following the topological limitations of the road. Each vehicle can connect to the servers at the fog nodes or CDC via wireless service offered by an MNO. Each vehicle $k\in\left\{1\ldots,K \right\}$ can generate a sequence of state updates to be uploaded and processed by servers at fog nodes or CDC and wait for the response/calculation results from the selected servers.
	\item[2)] Fog nodes: a set of mini-servers deployed close to the end users (e.g. vehicles) and can be accessed via a commercial wireless networks, e.g., a LTE network. Fog computing networks have typically been carefully planned by the MNO and each fog node can only serve end users within an exclusive area. Each fog node has a limited computational capacity and can only serve a limited number of vehicles at the same time.
	\item[3)] Cloud data center (CDC): CDC consists of high-performance servers that can offer high speed computational services at an affordable price, especially compared to the fog computing network. CDC typically provides service through the Internet and can serve vehicles located in a much wider service area than a fog node. However, It is often deployed in remote areas and therefore has much higher communication latency compared to the fog computing networks.
\end{itemize}

\section{AoI for Cloud/Fog Computing-Supported System}
\vspace{-0.12cm}
In this paper, we focus on optimizing the AoI of a cloud/fog computing-supported vehicular system. Let $G_0,G_1,\ldots$ be the time points for the vehicle to generate and send its state update. Each vehicle will send its update to the intended server immediately after generation. Once the update is received, fog nodes and CDC will process the update time duration, $\tau ^f$ and $\tau^c$, respectively before sending back the processed result to the vehicle.
We can then write the total service time of update $i$ from the update generated by a vehicle till the reception of the processed result as $T^c_i=R^c_i+\tau^c$ and $T^f_i=R^f_i+\tau^f$ for CDC and fog nodes, respectively, where $R^c_i$ and $R^f_i$ are round-trip time (RTT) for sending update and receiving result to and from CDC and fog nodes respectively. The result reception time points for the $i$-th update processed by fog node and CDC are given by $D^f_i=G_i+T_i^f$ and $D^c_i=G_i+T_i^c$, respectively. We follow the same line as  \cite{5984917} and define the AoI as the freshness of the most recently processed result received by the vehicle. More specifically, suppose the first update signal 0 is generated by the vehicle at time $G_0=-T_0$ and the corresponding result will be processed and received at $D^f_0=0$. The most recently received result by the vehicle in time $t$ will be time-stamped at $W(t)=\max\left\{G_i:D_i\leq t \right\}$, where
$$D_i=
\begin{cases}
D^f_i,&\text{if the $i$th update is processed by fog server};\\
D^c_i,&\text{otherwise}.
\end{cases}$$
We can then define AoI as:
\begin{equation}
A(t)=t-W(t).
\end{equation}

It can be observed that the AoI is a stochastic process that is closely related to the signal generation frequency, round-trip-time, processing delay, driving behavior, and the choice of cloud/fog servers. In this paper, we focus on empirical analysis and optimizing AoI via path planning for a cloud/fog-supported connected vehicular system. In fact, we observed that due to the random nature of wireless connection, AoI of a connected vehicle may exhibit significant temporal and spatial fluctuations. In other words, instead of optimizing the instantaneous AoI at each given time and location, it is more realistic to optimize the confidence/probability that a certain AoI requirement can be met. Therefore, in this paper, we use confidence level, defined as the probability that the AoI experienced by a vehicle can be guaranteed to be no greater than a given value, as the main performance metric for a connected vehicular system. We assume each vehicle can only access either fog server or CDC server when driving to a given location denoted as $l$. Let $x_k$ be the decision for vehicle $k$ to submit its workload to fog $(x_k=f)$ or CDC $(x_k=c)$. We define the confidence level of a connected vehicle $k$ when driving to location $l_k$ with the maximum tolerance level of AoI $\bar{A}$(ms) as:
\begin{equation}
C_k(l_k,x_k)=\Pr(A_k\leq \bar{A}).
\end{equation}
\subsection{Empirical Modeling and Analysis}
To evaluate the performance of AoI, we need to substitute our observed RTTs into (1) and generate the empirical AoI under different settings. According to (1), in addition to RTTs, AoI can be affected by three main factors:
\begin{itemize}
\item 1) Update generation frequency
\item 2) Workload processing delay at fog node and CDC
\item 3) Choice of servers (fog node or CDC).
\end{itemize}
We give a more detailed discussion as follows:
\\{\bf Update generation frequency}: Generally speaking, different connected vehicular services could require different update generation frequencies. For example, some infotainment services such as video streaming may require $24-30$ frames per second of updating frequency while highway road warning service requested an update every 2 seconds. In Fig. 1(b), we present the empirical PDFs of AoI under different update generation frequencies. We can observe that the variance of AoIs increases with the update generation frequencies. This is because the AoI only counts the most recent results received by the vehicle and if the processing result of the next update is received earlier than the result of the previous update, the previous result will be ignored. In other words, AoI ignores the result when the RTT plus the processing delay is much higher than the update generation interval.
\\{\bf Process Delay}: To evaluate the impact of processing delay on the AoI, we present the confidence level of AoI with update generation interval at 20ms under different services latency requirements. We can observe that the confidence level drops substantially when the processing delay plus the RTT exceeds the maximum tolerable latency of each supported service.
\\{\bf Choice of cloud/fog servers}: To evaluate the impact of the choice of cloud/fog servers on the AoI, we consider two scenarios as follows:
\begin{itemize}
\item Low Update Generation Frequency ($D^c_i<G_{i+1}$): In this case, choosing cloud and fog servers will result in different AoI updating frequency. The difference between AoI offered by fog nodes and AoI offered by CDC corresponds to the size of shadowed parallelogram shown in Fig. 2(a).
\item High Update Generation Frequency ($G_{i+1}<D^c_i<D^c_{i+1}$): There are two possible cases if the next updated signal is generated before reception of the processed result of the current update. If choosing the servers at fog or CDC for processing the update $i+1$ does not offer the sequence of signal updates, i.e., $D^c_i<D^f_{i+1}$, difference between AoI for choosing fog and cloud servers will be the same as that of Low update generation frequency scenario shown in the shadowed parallelogram in the left most triangle in Fig. 2(b). However, if switching from the cloud server to fog server causes update $i+1$ being received earlier than the feedback result of update $i$, the AoI difference caused by the vehicle to switch between fog and cloud server will result in two parallelogram areas in AoI. This is because if update $i+1$ can be received earlier than update $i$, the AoI will no longer be affected by update $i$, resulting in two parts of reduction caused by latency reduction for choosing fog and cloud server to process update $i+1$. 
\end{itemize}
\begin{figure}[htbp]
	\centering	
	\subfigure[]{
		\begin{minipage}[t]{0.45\linewidth}
			\centering
			\includegraphics[width=1.7in]{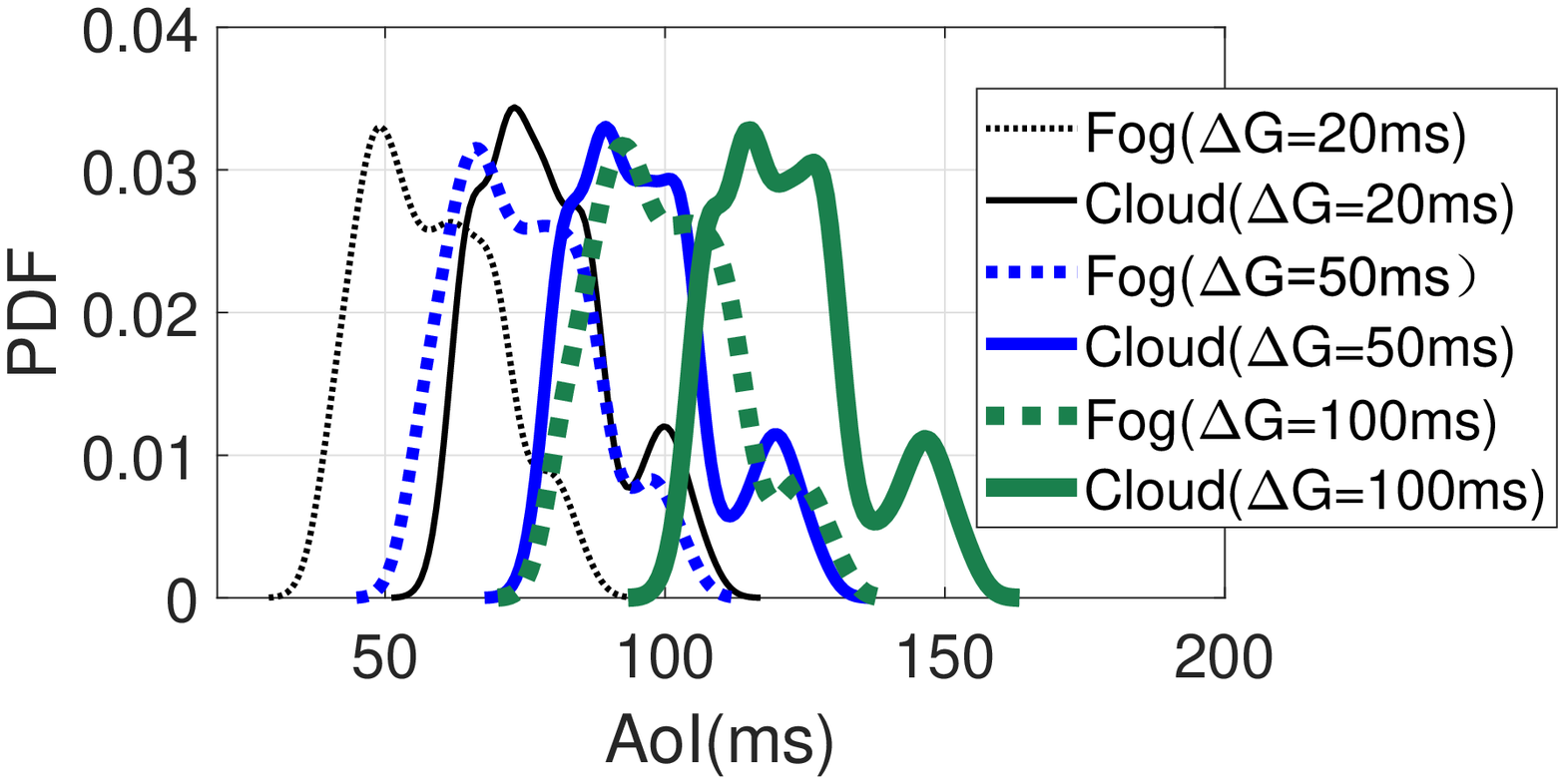}
		\end{minipage}%
	}
	\subfigure[]{
		\begin{minipage}[t]{0.45\linewidth}
			\centering
			\includegraphics[width=1.7in]{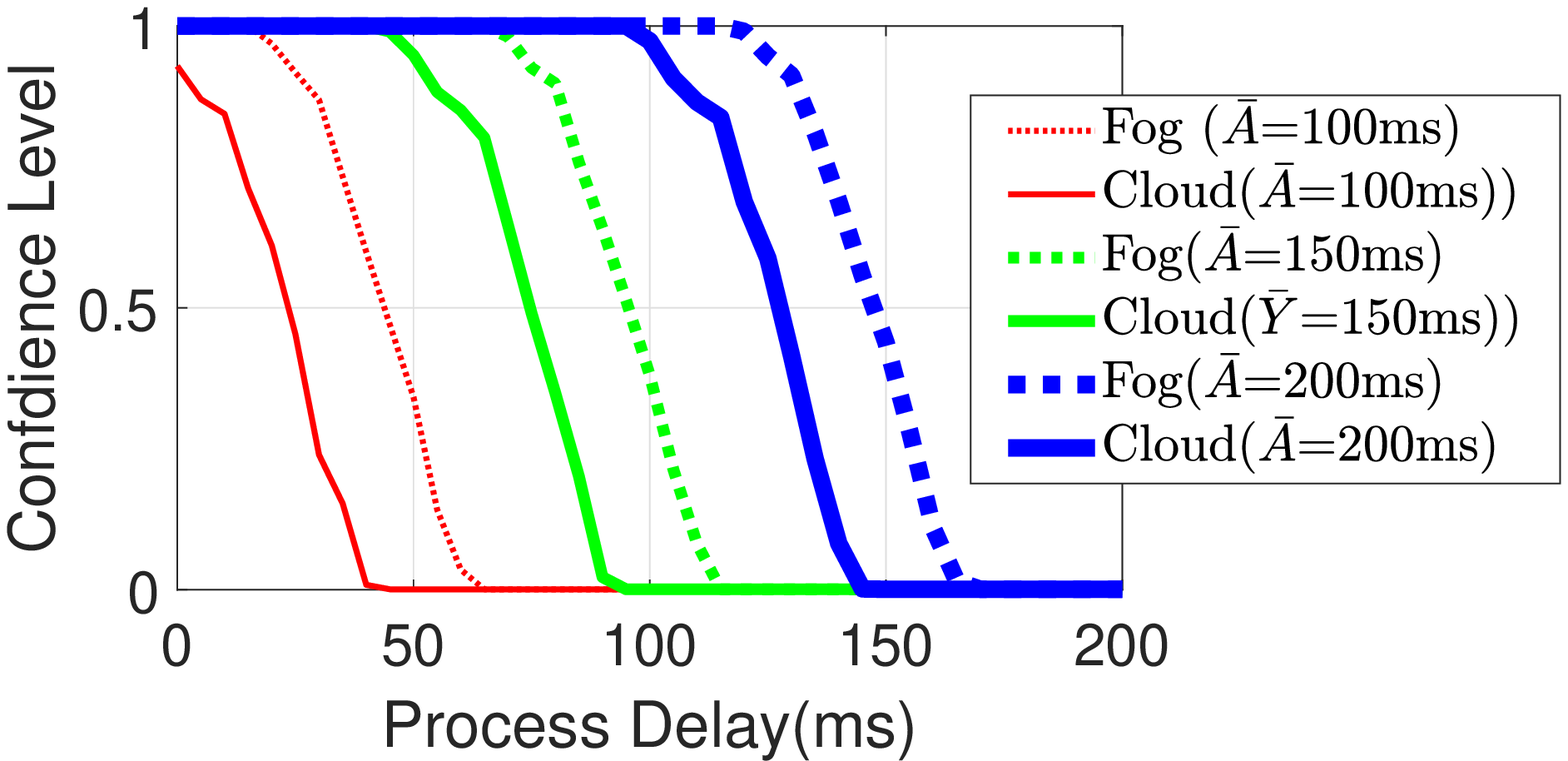}
		\end{minipage}
	}
	\centering
	\vspace{-0.12in}
	\caption{(a) PDFs of AoI under three different $\bar A$ in fog/cloud changing as the generation interval increasing (suppose the process delay can be ignored). (b) CDFs of AoI under three different $\bar A$ in fog/cloud changing as the process delay increasing. The generation interval is chosen as $\Delta G=20ms$.}
\end{figure}
\begin{figure}[htbp]
	\centering
	\subfigure[]{
		\begin{minipage}[t]{0.4\linewidth}
			\centering
			\includegraphics[width=1.6in]{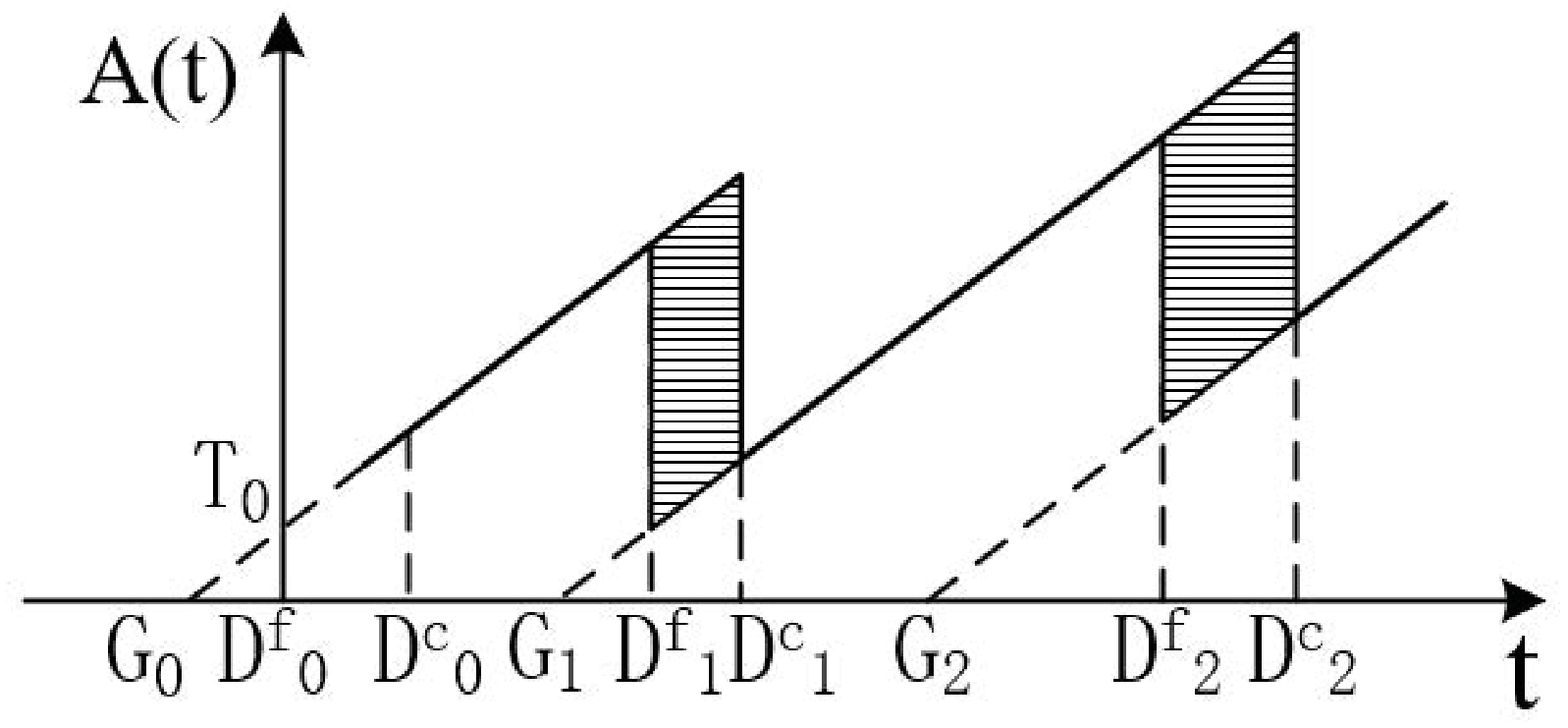}
		\end{minipage}%
	}%
	\subfigure[]{
		\begin{minipage}[t]{0.4\linewidth}
			\centering
			\includegraphics[width=1.6in]{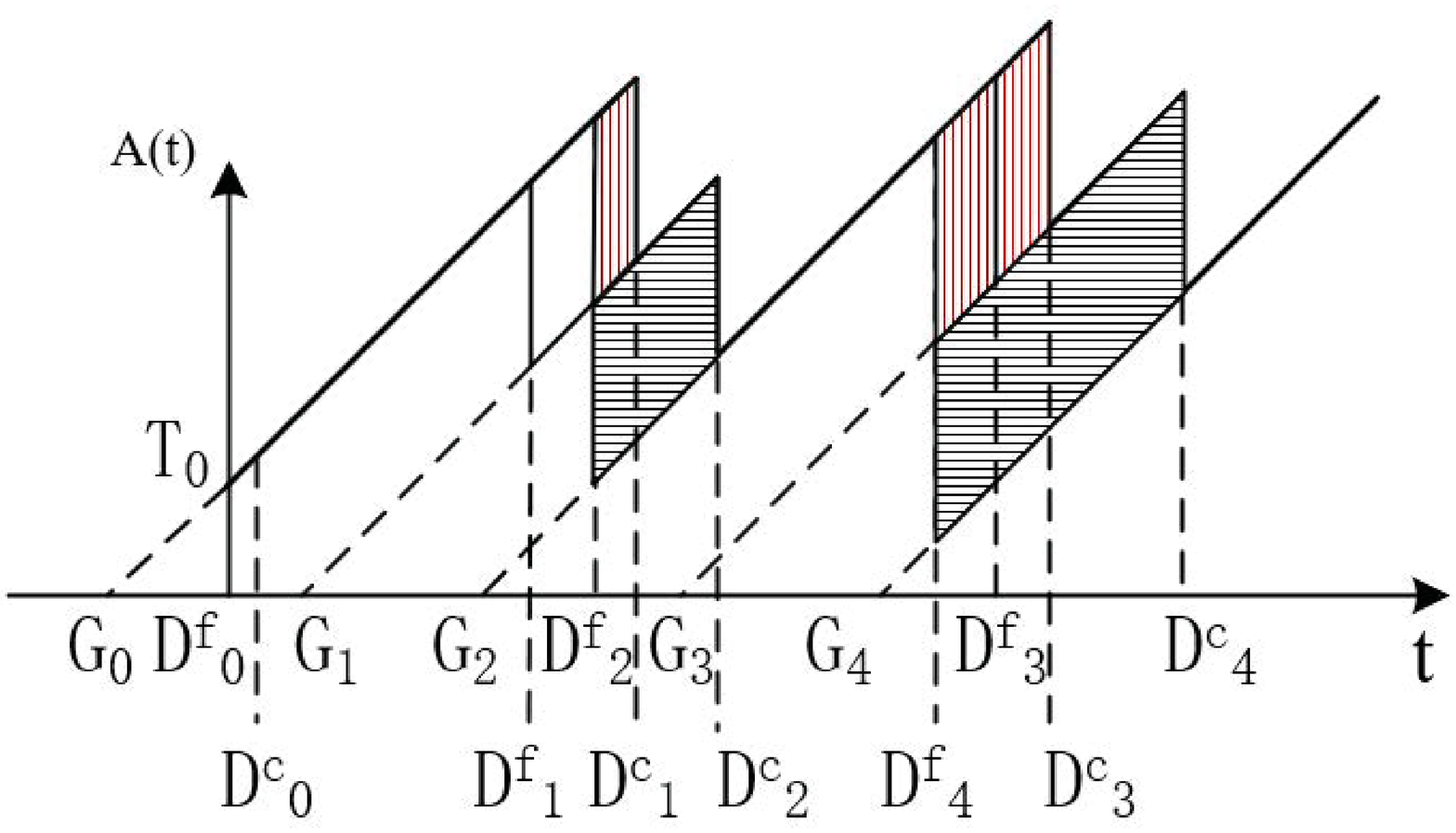}
		\end{minipage}%
	}%
	\centering
	\vspace{-0.12in}
	\caption{(a) The variation of AoI when switching from cloud to fog under the condition that update generation frequency is low. (b) The variation of AoI when switching from cloud to fog under the condition that update generation frequency is high.}
\end{figure}
\subsection{Measurement Campaign}
As observed in (1), the AoI of a connected vehicle is closely related to back-and-forth communication delay for workload submission and result feedback. We therefore begin by collecting traces to evaluate the communication latency between multiple connected vehicles and the fog/cloud servers. One straightforward way to measure the back-and-forth communication latency is to test the RTT between vehicles and networks. We develop an Android App to periodically ping different networks and record the RTT. Motivated by the 5GAA whitepaper \cite{5GAAUseCases} that most safety-related vehicular applications require updating information with a limited packet size smaller than 996 byte, we set the size of data package in ping command as 996 byte. Except for the RTT, our developed APP also records other vehicle-related information such as location, speed, altitude and connected network as well. To evaluate RTT between vehicles and the fog server, we follow the same line as  \cite{tordera2016fog} and assume the IP address of service-gateway (S-GW) as the fog node server location. For the RTT between vehicles and the CDC, we use the IP address of a cloud server deployed by a major cloud service provider in China. In order to measure the RTT in different situations, 6 smart phones installed with our APP is mounted on 6 campus shuttles for three months of measurement and over 1,000,000 traces were collected when driving throughout the campus.
\subsection{Empirical PDF Classification \& Region Segmentation}
It has been observed that although the instantaneous value of RTT varies significantly between consecutive time slots and neighboring location points, the empirical PDFs of RTTs show strong temporal and spatial correlation. Due to the close relationship between AoI and RTT, the empirical PDF of AoI should exhibit similar correlations at different time across various locations. In particular, in Fig. 3(a), we present the average AoIs measured at different location points under a given source generating frequency. We can observe that the statistic feature of AoI measured at neighboring locations shows strong correlations. In addition, the empirical PDFs can be classified into a limited number of clusters. Vehicles driving within each region will experience similar latency PDF.

More formally, we can extend the K-means-based clustering approach by adopting the distance function of PDFs. In this paper, we follow a commonly adopted setting and use K-R distance $O(F,G)$ to measure the distance between two empirical PDFs $F$ and $G$ measured at two different regions, i.e., we have
\begin{equation}
O(F,G)=\int_{0}^{\infty}|F(t)-G(t)|\,dt.\\
\end{equation}\\
We present the detailed region segmentation algorithm in Algorithm 1.
\begin{algorithm}
	\footnotesize
	\caption{}\label{Algorithm 1}
	\begin{itemize}
		\item[] Set the Number of clustering centers as $R>1$.
		\item[] Initialization:
		\begin{itemize}
			\item[] 1. Selecting a distribution function sample randomly as the first clustering center.
			\item[] 2. Select the remaining K-1 clustering centers.\\
			{\bf while} $r \leq R-1$ {\bf do}
			\begin{itemize}
				\item[] 1) {\bf For} $i=1$ {\bf to} $P$ {\bf do}
				\begin{itemize}
					\item[] Calculate the KR distance from sample $i$ to $(R-r)$ centers and record the minimum distance as $d_i$.
					\item[] {\bf End For}
				\end{itemize}
				\item[] 2) Find the maximum $d_i$ and choose sample $i$ as the next
				clustering center.
				\item[] 3) $r=r+1$.
			\end{itemize}
			{\bf End while}
		\end{itemize}
		\item[] Set the maximum number of iterations as $M>0$
		\item[] {\bf while} $m \leq M$ {\bf do}
		\begin{itemize}
			\item[] 1. {\bf For} $i=1$ {\bf to} $P$ {\bf do}
			\begin{itemize}
				\item[] Calculate the KR distance from sample $i$ to $R$ clustering centers and select the closest center as the category for sample $i$.\\
				{\bf End For}
			\end{itemize}
			\item[] 2. Calculate the mean value for each category and replace the cluster center with the mean value.
			\item[] 3. $m=m+1$.
		\end{itemize}
		{\bf End while}
	\end{itemize}
\end{algorithm}

\begin{figure}[htbp]
	\centering
	\subfigure[]{
		\begin{minipage}[t]{0.5\linewidth}
			\centering
			\includegraphics[width=1.7in]{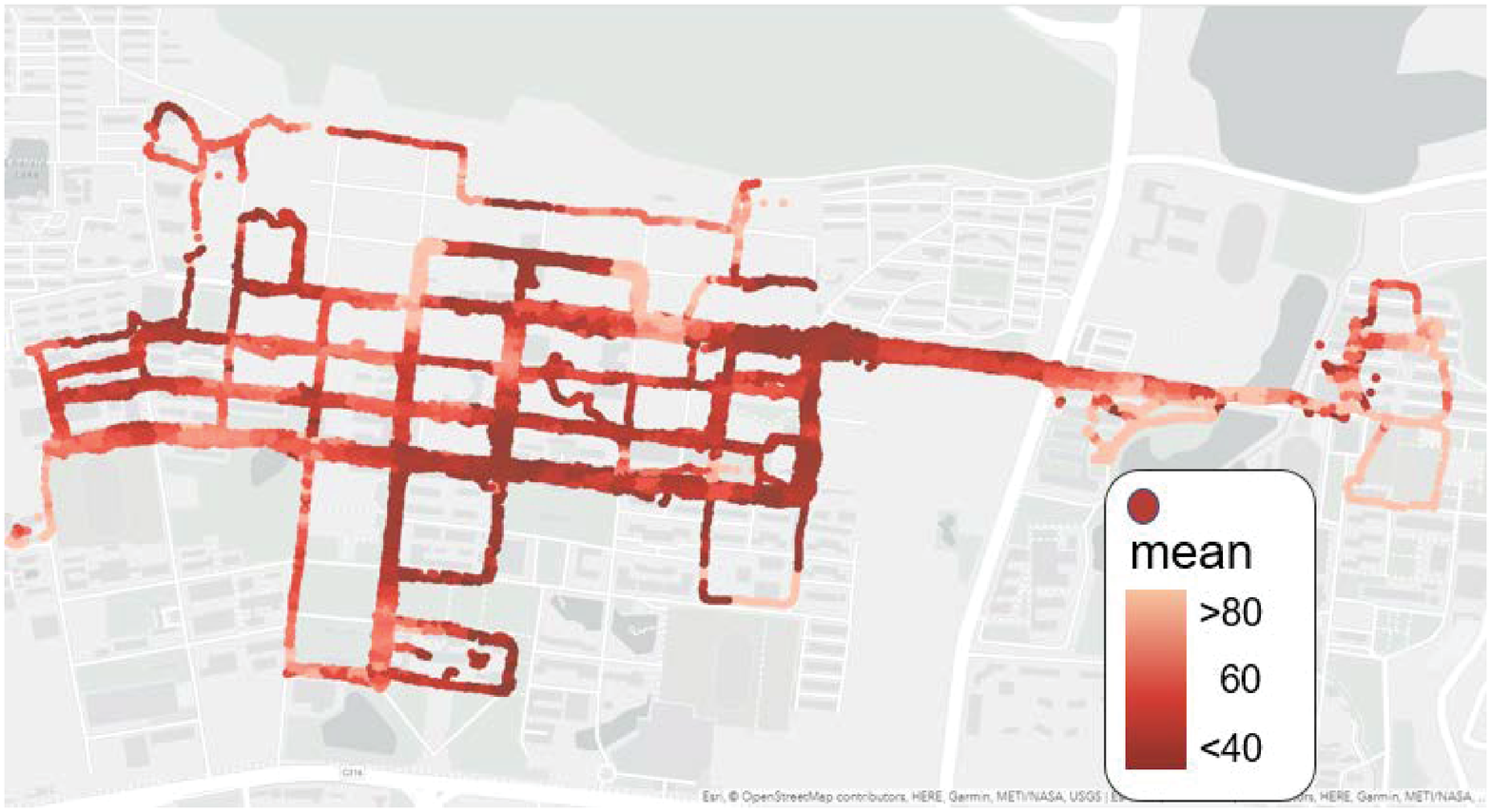}
		\end{minipage}%
	}%
	\subfigure[]{
		\begin{minipage}[t]{0.5\linewidth}
			\centering
			\includegraphics[width=1.5in]{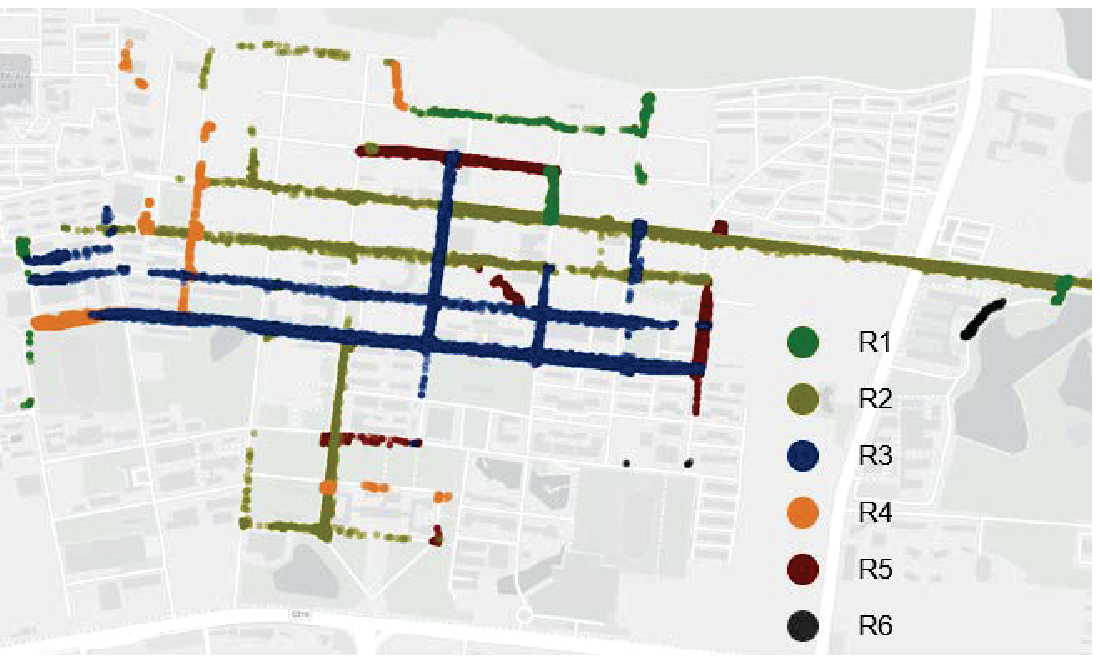}
		\end{minipage}%
	}%
	\centering
	\vspace{-0.12in}
	\caption{(a) Average AoIs measured at different location points under a given source generate frequency $\Delta G=20ms$. (b) Result of region segmentation.}
\end{figure}
\section{DQN for multi-vehicles path planning}
In this section, we propose a novel architecture of multi-vehicle path planning. 
We consider confidence of AoI and focus on maximizing  the guaranteed probability that a certain AoI constraint can be satisfied when multiple connected vehicles are driving within a given area. 
\subsection{Problem formulation}
Consider a vehicular  network with multiple vehicles driving within the same location 
covered by a set of fog nodes and a remote CDC. We assume each vehicle  is driving from a given starting location and can choose different routes to its intended destination. 
We consider a slotted process and assume the driving behavior of each vehicle can be considered as a Markov decision process (MDP) with finite horizon, in which its current location and AoI performance only depends on its location in the previous time slot. 
\begin{itemize}
\item State space ${\cal S}$: Suppose the vehicle can dynamically switching between different driving routes at each intersection. We define the state as a finite set of all the possible routes between the source and destination. Each instance of state in time slot $j$ as $s_j \in {\cal S}$ can be written as $s_j=\langle l_j\rangle.$
%
\item Action space ${\cal A}$: is a finite set of all the possible direction choices for the vehicle at each intersection, such as right and left. We write the action choice in time slot $j$ as $a_j \in {\cal A}_{s_j}$ for all $j$. 
\item State Transition Function ${\cal T}$: ${\cal S} \times {\cal A} \times {\cal S}\rightarrow[0,1]:$ The choice of action determines the next state with certainty. We define a mapping $f(s,a)$, from $S\times A_s$ to $S$. Then we can write the probability of state transferred from state $s_j$ to $s_{j+1}$ when taking action $a_j$ as:
\begin{equation}
\mbox {Pr}(s_{j+1}|s_j,a_j)=\left\{
\begin{aligned}
1,\ &f(s_j,a_j)=s_{j+1};\\
0,\ &\mbox{otherwise}.\\
\end{aligned}
\right.
\end{equation}
\item Utility Function: The main objective is to maximize the general confidence of AoI during the driving process. We define the AoI confidence during the process from time slot $j$ to time slot $j+1$ as $r_j$. 
Suppose the vehicle getting to the destination at time slot $N$. Then we calculate the general AoI confidence as:
\\\begin{equation}
U=\prod_{i=1}^{N-1}{r_i}.
\label{eq_Utility}
\end{equation}
\end{itemize}

To maximize the long-term reward, the current and future payoff must be jointly considered. We define the value function $Q(s_j,a_j)$ when action $a_j$ is taken at state $s_j$ as
\\\begin{equation}
Q(s_j,a_j)=\left\{
\begin{aligned}
&r_j,\ {\mbox {if}}\ s_{j+1}\ {\mbox {is terminated}};\\
&r_j*\max\limits_{a\in {\cal A}_{s_{j+1}}}{Q(s_{j+1},a)}^\gamma,\ \mbox{otherwise}.
\end{aligned}
\right.
\end{equation}
We define the optimal value function for state $s_j$ as
\\\begin{equation}
Q^{*}(s_j)=\max\limits_{a_j\in {\cal A}_{s_j}}Q(s_j,a_j).
\end{equation}

Hence, the optimal policy $\pi^*$ is given by
\\\begin{equation}
\pi^*= \arg\max\limits_{a_j\in {\cal A}_{s_j}}Q(s_j,a_j).
\end{equation}

\subsection{Deep Q-learning}
Q-learning is a effective approach for solving MDP problem when the sizes of state and action spaces are small. However, for problem with a large number of states and actions, the convergence rate will become slower\cite{szepesvari1998asymptotic}. In this paper, we focus on path planning problem, where the state space is large. 
We adopt deep Q-learning approach proposed by Deepmind\cite{van2016deep} 
to find the optimal solution with large state-space. 

DQN replaces the Q-table with a deep neural network which can be used to estimate the value of $\langle s_j,a_j\rangle$. 
DQN only updates the parameter of neural network instead of updating the value of each pair of $\langle s_j,a_j\rangle$, which can take much less time cost than Q-learning especially for problems with large state and action space. Normally, DQN consists of  four parts, i.e., feature input, experience replay pool, predict network and target network.
\begin{itemize}
	\item Feature set: Feature is the input of neural network. In this problem, we determine the routes between the source and destination as the feature. 
	\item Experience replay pool: The transition $(s_j,a_j,r_j,s_{j+1})$ observed after each performance is stored in the experience replay pool. The labeled sample for learning process is selected from the experience replay pool. The dependency of samples are eliminated by applying stochastic method.
	\item Predict network: For each state, the algorithm selects an action through the predict network. The predict network evaluates the value after each possible action, then the best action based on predicting result will be taken with possibility $1-\varepsilon$.
	\item Target network: After state transition, the value of the next state is evaluated by the target network. In order to improve the stabilization of algorithm, the parameter of target network updates to the predict network slowly, which eliminates the dependency between choosing action and evaluating value and reduces the network vibration.
\end{itemize}
\begin{figure}[ht]
	\centering
	\includegraphics[scale=0.25]{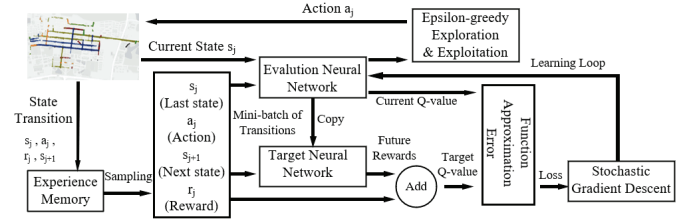}
	\vspace{-0.12in}
	\caption{DQN structure}
\end{figure}
The structure of DQN is given in Fig.4, and more details are shown in Algorithm 2.
\begin{algorithm}
	\footnotesize
	\caption{}\label{Algorithm 2}
	\begin{itemize}
		\item Initialization:
		\begin{itemize}
			\item[1.] Initialize replay memory to capacity $D$
			\item[2.] Initialize predict network with random weights $\theta$
			\item[3.] Initialize target network with weight $\theta^-=\theta$
		\end{itemize}
	    \item Iteration:\\
	    {\bf for} episode 1 to T {\bf do}
	    \begin{itemize}
	    	\item[1.] Repeat:
	    	\begin{itemize}
	    		\item[1)] With probability $\varepsilon$ choose a random action $a_j$ otherwise select $a_j=\max\limits_{a \in{\cal A}_{s_j}}Q^*(s_j,a;\theta)$
	    		\item[2)] Perform action $a_j$ in emulator and observe reward $r_j$ and next state $s_{j+1}$
	    		\item[3)] Store transition $(s_j,a_j,r_j,s_{j+1})$ in the replay memory
	    	\end{itemize}
            \item[] Until $s_{j+1}$ terminal
	    	\item[2.] {\bf If} replay memory pool is full: \\
	    	Sample random minibatch of transitions $(s_j,a_j,r_j,s_{j+1})$ from the replay memory pool\\
	        {\bf else}: \\
	    	continue
	    	\item[3.] $y_j=r_j*\max\limits_{a' \in {\cal A}_{s_j+1}}Q(s_{j+1},a';\theta^-)^\gamma$
	    	\item[4.] Perform a gradient descent step on $(y_j-Q(s_{j},a_{j};\theta))^2$
	    	\item[5.] Every C steps reset $\theta^-=\theta$
	    \end{itemize}
      {\bf end for}
	\end{itemize}
\end{algorithm}
\subsection{Simulation result and analysis}
\begin{figure}[htbp]
	\centering
	\subfigure[]{
		\begin{minipage}[t]{0.5\linewidth}
			\centering
			\includegraphics[width=1.75in]{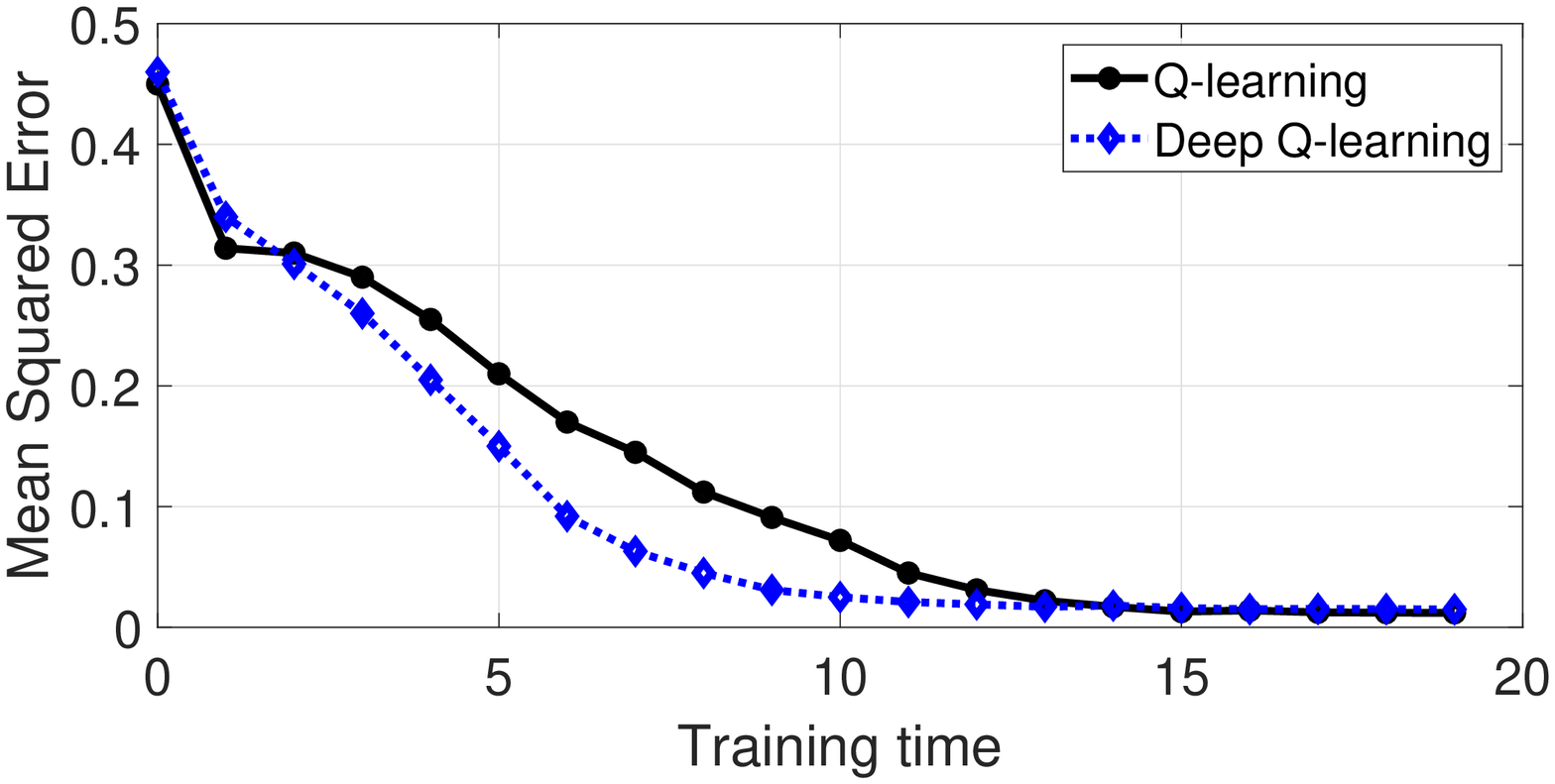}
		\end{minipage}%
	}%
	\subfigure[]{
		\begin{minipage}[t]{0.5\linewidth}
			\centering
			\includegraphics[width=1.75in]{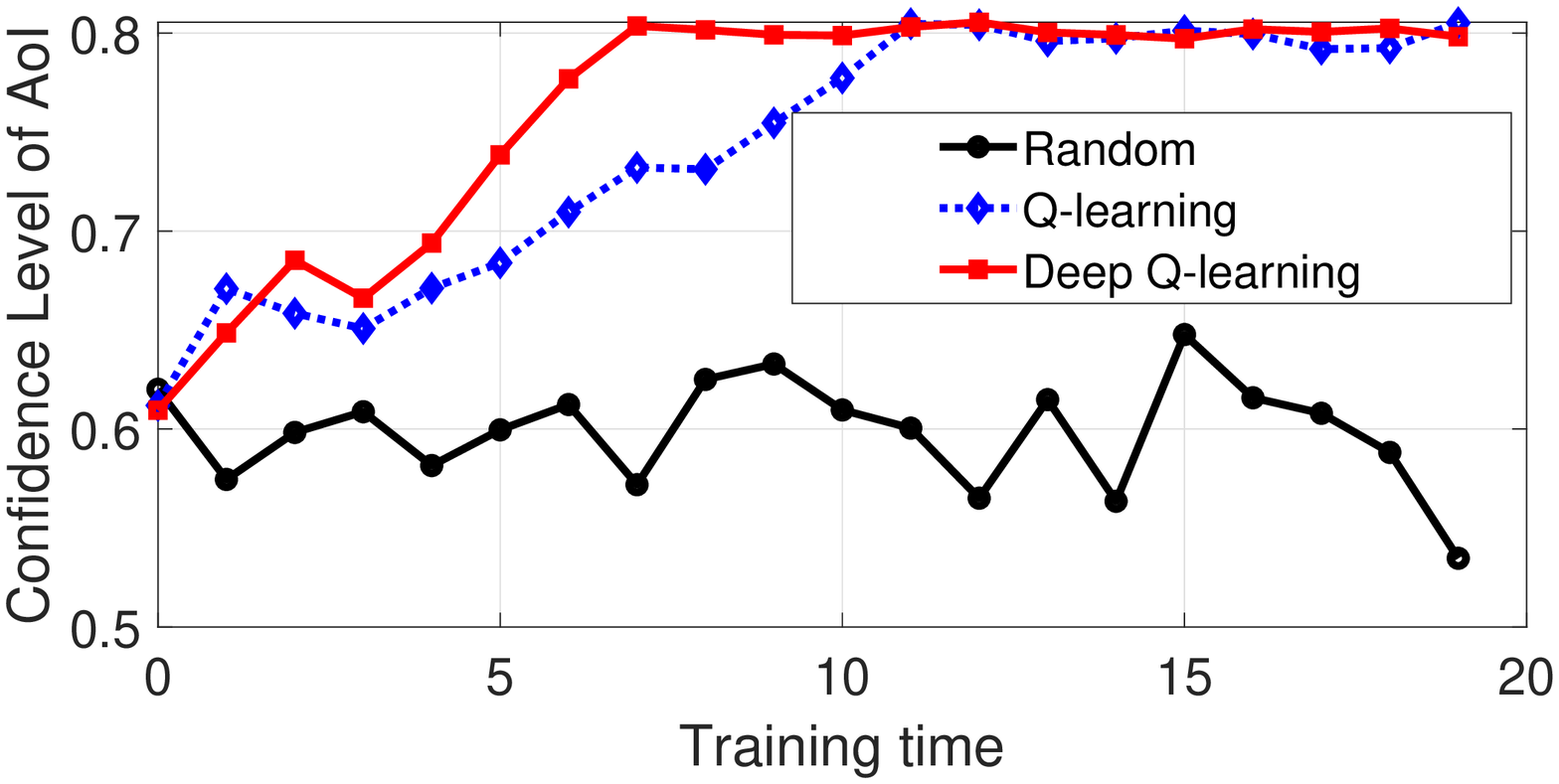}
		\end{minipage}%
	}%
	\centering
	\vspace{-0.05in}
	\caption{(a) Mean Squared Error of DQN compared with Q-learning. (b) Average confidence of AoI running in DQN and Q-learning compared to the random policy.}
\end{figure}

In Figure 5, we compare the convergence performance of Q-learning and DQN when being applied to maximize the long-term confidence level of AoI for a connected vehicle using the setting described in the previous subsections. We can observe  that both Q-learning and DQN are able to reduce the driving cost and improve the AoI performance. 
In Figure 5(a), we observe that compared to Q-learning, DQN converge in a faster speed and can approach to a minimized error in around 10 training times. To evaluate the improvement that can be achieved by different policies with different training time, we present the AoI confidence defined in (\ref{eq_Utility}) for various driving routes selected by the vehicle using different policies with different training time. We can observe that DQN converges to the highest overall confidence of AoI after two to three times of training.   
\section{Conclusion}
This paper studies the AoI for fog/cloud-supported vehicular systems. We report an empirical study of AoI on a multi-vehicular campus shuttle system. The impact of update generation frequency, processing delay, and choice of fog/cloud servers on the confidence of AoI have been investigated. Motivated by the observation that the empirical PDF of AoI exhibits strong spatial correlation, we propose a modified K-mean-based clustering approach to categorize the empirical PDF at different location points throughout our considered area into a limited number of probability distributions. Finally, we investigate driving route planning with the main objective to optimize the confidence of AoI. A DQN-based approach is introduced to find the optimal driving policy. Numerical results show that our proposed algorithm can significantly reduce the driving cost and improve the average AoI performance.

\section*{Acknowledgment}
The authors would like to thank Ericsson (China) Hubei Branch and  China Mobile Hubei 5G Joint-innovation Lab for help in the data collection.

\vspace{-0.12in}
\bibliographystyle{IEEEtran}
\bibliography{2020ICC}
\end{document}